\documentclass[10pt,a4paper]{article}
\usepackage{graphicx}
\usepackage{tabularx}
\usepackage{multicol}
\usepackage{tabto}
\usepackage{url}
\usepackage{enumitem}
\usepackage{textcomp}
\usepackage{eurosym}
\usepackage{wasysym}
\usepackage{xcolor}
\usepackage[labelfont=bf]{caption}
\usepackage[yyyymmdd]{datetime}

\usepackage{flowchart}
\usetikzlibrary{shapes,arrows.meta,chains,external}

\begin{document}

\bibliographystyle{plain}
\thispagestyle{empty}

\begin{center}
\large{\bf Privacy by Design in Value-Exchange Systems}\\
\end{center}
\vspace{0.5em}
\begin{center}
{\large Geoffrey Goodell}\\
\vspace{0.5em}
{\large University College London}\\
\vspace{0.5em}
{\large \texttt{g.goodell@ucl.ac.uk}}
\end{center}

\begin{center}
{\textit{This Version: \today}}\\
\end{center}
\vspace{4em}

\begin{quote}

``The cashless and checkless society is obviously technologically feasible now
and will soon be economically feasible. (Political feasibility is another
question.) Personally, the thought of a system that will record and store
information on what I purchased for how much at what time and place everytime I
purchase so much as a newspaper or candy bar frightens me.  And that brings me
to the privacy question.''  \textit{--- Paul Armer, the RAND Corporation,
1967~\cite{armer1967}}

\end{quote}

\vspace{4em}

\noindent This article addresses some of the most contentious issues related to
privacy in electronic payment systems, particularly the current zeitgeist of
proposed solutions for central bank digital currency, in support of arguments
made by Goodell, Al-Nakib, and Tasca~\cite{gnt2020}.

\section{Core Principles}

We begin with some terms.  We use the term \textit{privacy} to characterise a
situation in which a person is not forced to reveal his or her attributes or
transactions in a manner that allows those attributes or transactions to be
linkable to each other or to the person.  This contrasts with \textit{data
protection}, which concerns the unauthorised use of data containing such
linkages once they have already been revealed.  In the context of a system,
therefore, privacy is an architectural characteristic of the design of a system
wherein its users are not forced to reveal linkages among their transactions or
attributes in the course of using the system.  We refer to this architectural
characteristic of systems as \textit{privacy by design}.

We use the term \textit{exceptional access mechanism} to refer to an element of
system design wherein a privileged authority (or other actor, perhaps a hacker
or insider) can access data linking the attributes or transactions of an
individual.  Systems containing exceptional access mechanisms are never private
by design, not because of the specific mechanisms for accessing the linkages
but because the systems force their users to reveal the data in the first
instance.  Furthermore, exceptional access mechanisms are widely understood to
introduce security vulnerabilities and increase the scope for
abuse~\cite{abelson1997,abelson2015,benaloh2018}.  We specifically note that
privacy cannot be ``provided'' or ``guaranteed'' by any particular authority or
third-party.

We use the term \textit{value-exchange system} we refer to a system that people
can use to exchange value, including but not limited to payment methods and
networks in which people exchange goods and services for money.

We believe that ordinary people conducting ordinary, legitimate activities, for
everyday purposes should use value-exchange systems that are private by design
to conduct those activities.  This is necessary to ensure that individuals can
conduct their ordinary, private business without fear that their behaviour will
be profiled.  Specifically, we mean to imply that transactions conducted using
a value-exchange system that is private by design should not intrinsically be
linkable to information about both counterparties.

\section{Money Laundering}

We begin by asserting that AML/KYC regulations around the world, including but
not limited to rules intended to achieve conformity with the FATF
recommendations~\cite{fatf-recommendations}, should not be construed as a broad
admonition against private exchange of value.  First, such regulations apply to
a particular set of businesses and institutions that handle money on behalf of
others.  They do not directly apply to all users of money.  Second, such
regulations were established at a point in history when, unlike today, ordinary
people had the option to use cash, a broadly anonymous mechanism for exchanging
value, as an instrument in wide range of ordinary transactions, and when the
manifest ability for powerful state and non-state actors to collect, aggregate,
and analyse data to build profiles of individuals throughout entire populations
was far smaller.  It is not because of AML/KYC regulations that ``surveillance
capitalism''~\cite{zuboff2015} has become a leviathan, but the fact that the
world has changed means that we need to rethink the contours and limits to what
such regulations should achieve.

By stating our requirement for privacy by design, we do not mean to imply that
value-exchange systems must allow peer-to-peer transactions without the
involvement of regulated intermediaries.  Instead, we mean to imply that
individuals must be allowed to transact without the need to trust any third
party to ensure that they will not be profiled.  To satisfy this requirement,
it would be sufficient for regulators to have visibility into every
transaction, provided that there is no expectation that they will have access
to information that could identify both counterparties.  We argue that a
transaction to or from an anonymous user should be allowable under ordinary
circumstances, provided that regulators can identify one of the counterparties
to every payment (generally, the recipient), and with the expectation that
regulated institutions would share whatever data are needed to satisfy AML/KYC
regulations.  We note that regulated intermediaries might have never received
the counterparty information in such cases.

Following the argument from Goodell, Al-Nakib, and Tasca~\cite{gnt2020},
businesses do not generally need to know who their customers are, and their
banks certainly do not need to know who the customers of their customers are.
Cases wherein businesses need to know their customers can be handled outside
the context of the payment mechanism, and cases wherein businesses are
concerned about assurance of payments can be handled by escrow services.
Additionally, payments for extraordinary or large-value goods and services can
be handled on an exceptional basis without privacy by design, and such
requirements can be made legally binding for that subset of payments without
providing exceptional access to all payments.

\section{Tax Evasion}

In most countries that conform to IFRS or GAAP accounting standards, government
authorities are assumed to have a legitimate interest in knowing and auditing
the income of all individuals, partnerships, and corporations.  We interpret
this to mean that for transactions of value that can be interpreted to
constitute a payment, authorities have a legitimate interest to associate each
transaction with its recipient.  Note that this does not mean that authorities
have a legitimate interest to know both the sender and the recipient.

Therefore, we suggest that regulated intermediaries would generally require the
recipient of payments to be financial institutions, for the benefit of
particular account-holders of that institution.  Recipients of payments without
a bank account could opt to receive cash or digital tokens via a regulated
intermediary that might be required to conduct an identification procedure for
AML/KYC purposes and flag the relevant tax authorities for reconciliation, with
appropriate exceptions made for reimbursements and gifts.

In this sense, all remittances using a digital payment system could be
monitored, avoiding the peer-to-peer transactions that are viewed by
authorities as undesirable from the perspective of taxation and the social
contract without creating infrastructure that could be used to conduct mass
surveillance by linking recipients of payments to their counterparties.

We also suggest that payers could obtain attribute-backed receipts from their
counterparties that could be used as evidence of eligibility for tax deductions
or credits.  Claimants would be encouraged but not required to file such
claims, and incentives would be aligned such that, all else being equal, they
would be inclined to do so.  Using zero-knowledge proofs, such receipts could
even be reconciled without linking specific receipts to specific claims.

\section{Common Fraud}

There are many forms of fraud, and we specifically argue that it is better to
empower ordinary citizens to detect fraud than to create exceptional access
mechanisms for authorities.  We quote directly from Goodell, Al-Nakib, and
Tasca:

\begin{quote}
``Because our system allows a measure of true anonymity, it does not
intrinsically require the identities of both counterparties to be visible to
authorities.  Revealing mutual counterparty information for every transaction
would divert the onus of fraud detection to law enforcement agencies,
effectively increasing their burden, while well-motivated criminals would still
be able to use proxies or compromised accounts to achieve their objectives,
even if every transaction were fully transparent.''~\cite{gnt2020}

``Our system design offers a different approach.  Because every transaction
involves a regulated financial intermediary that would presumably be bound by
AML/KYC regulations, there is a clear path to investigating every transaction.
Authorities would be positioned to ensure that holders of accounts that take
payments from private wallets adhere to certain rules and restrictions,
including but not limited to tax monitoring.  The records from such accounts,
combined with the auditable ledger entries generated by the DLT system, could
enable real-time collection of data concerning taxable income that could
support reconciliation and compliance efforts.  Because all of the retail
payments involving digital currency would use the same ledger, identification
of anomalous behaviour, such as a merchant supplying an invalid destination
account for remittances from private wallets, would be more straightforward
than in the current system, and real-time automated compliance would be more
readily achievable.  Such detection could even be done in real time not only by
authorities but also by counterparties, thus reducing the likelihood that it
would occur in the first instance.''~\cite{gnt2020}
\end{quote}

\section{System Operation}

With the focus on privacy, it might be reasonable to ask: Why do we need the
transactions to be processed by private actors, who have obvious incentives to
capitalise on data that can be used to profile users, rather than by government
actors, who would at least be accountable to the public?  Were our approach
based on data protection, then we might indeed make such a recommendation.
However, privacy is not the same as data protection, and a system based on
privacy by design has a different set of design requirements.

First, we recognise that in a truly private system, the operators do not
actually have access to the information from which profiles might be
constructed in the first instance, because customers are not forced to reveal
it.  Thus, although the motivation and integrity of individual actors is
important, it is more important that customers have a wide range of independent
actors with which they can interact, so that none of them are in a position to
compel users to reveal ancillary metadata outside the system requirements that
could be used to build profiles.

Second, only in a federated system of private, independent actors can we
achieve a measure of assurance that the system operators are accountable to the
wider public.  Correct operation of the system is essential for a system that
is private by design, since the temptation to change the rules to undermine the
privacy of its users would surely be great.  We use the term
\textit{sousveillance} to refer to the means by which the governed can observe
the actions of those in power and ultimately hold them to account.  Because
proposed changes to the system must be circulated and agreed by members of a
federated system, any of whom could potentially object or even publicise news
of the proposed change, and because compliance with the rules can be observed
and audited, a federated system of private, independent actors offers a measure
of accountability to system operators.

Third, we recognise that a monopoly operator lacks an incentive to maintain
privacy characteristics through continued improvement.  Competition in general
and a diversity of implementations in particular offer scope for innovation
that will lead to more robust design characteristics and outcomes.

\section{Data Brokerage}

Privacy is a public good and susceptible to the Tragedy of the Commons.  We
believe that it is not enough to make it possible for users to make payments
without identifying themselves.  Unless a system is private by design and
private \textit{by default}, the anonymity set of users who need privacy the
most will be undermined by the preponderance of people who are not part of that
set.

The autonomy of individuals around the world is under threat every day by data
harvesters, data brokers, and data consumers who build profiles for the purpose
of manipulating populations cheaply and at scale.  With surveillance capitalism
in mind, we might imagine that regulations that mandate the implementation and
use of systems that are private by design could be a means by which state
actors could push back against powerful profilers in the interest of the
public.  The challenge of compelling powerful incumbents to change behaviour is
daunting, although there is evidence that the EU has recently begun to explore
ways to change the model that allows data brokers to be as successful as they
are~\cite{riley2019}.  Of course, the value of payments data is not lost on
global data brokers and profiling businesses, who are keen to access such
data~\cite{khan2019}.  Financial transactions are particularly sensitive as a
high-value source of information about the habits, predilections, and
circumstances of individual persons, not only because of the high degree of
assurance provided by AML/KYC procedures that are prerequisites to such
transactions in practice today but also because of the fact that exchanges of
value definitionally consume scarce resources that require judgement and
parsimony on the part of transacting parties.

In the meantime, full-service banks are being squeezed, on one side by
challenger banks that often have data-sharing relationships with data consumers
such as payment networks, retail marketplaces, credit bureaus, and insurance
firms~\cite{ghose2019}, and on the other side by costs associated with AML/KYC
compliance~\cite{lexisnexis}.  Although we do not suggest specific changes to
AML/KYC regulations, we do believe that a system that encourages banks to
collect less data rather than more about their customers' transactions would
benefit both the banks and their customers.

\section{Concluding Remarks}

Ultimately, the costs and benefits of dragnet surveillance are unevenly
distributed across the population.  Profiling, whether by state actors or
non-state actors, means that people at the margins face discrimination on the
basis of their status, while protections accrue to those who are already
well-endowed.  Furthermore, surveillance does not actually serve the purpose
for which it is intended, even if we ignore its costs to society.  Criminals
would not need a system that is private by design to achieve their objectives
because they will always have access to stolen credentials or account
information that will allow them to transact using the credentials of their
victims.  Individuals with the means to do so will also not need a system that
is private by design, as they always be able to rely upon members of their
retinues to conduct transactions on their behalf.  So really, the only people
that would be harmed by surveillance would be those who are both law-abiding
and relatively poor.  Thus, surveillance as a policy is intrinsically
regressive.

In conclusion, trust is something that must be earned.  Trust cannot be
compelled by a central operator, and that privacy must be intrinsic and not
subject to trust.  We must not countenance the deployment of a system that
demands that its users submit their everyday activities to monitoring and
profiling by powerful authorities.  Not only would such a system inexorably
create demand for alternative, ``outside solutions'' that would be accessible
only to a small segment of the population, but it would also undermine the
autonomy that is foundational to liberty and human flourishing.

\end{document}